\documentclass[conference]{IEEEtran}
\IEEEoverridecommandlockouts
\usepackage{cite}
\usepackage{amsmath,amssymb,amsfonts}
\usepackage{algorithmic}
\usepackage{graphicx}
\usepackage{subcaption}
\usepackage{textcomp}
\usepackage{xcolor}
\usepackage[export]{adjustbox}
\def\BibTeX{{\rm B\kern-.05em{\sc i\kern-.025em b}\kern-.08em
    T\kern-.1667em\lower.7ex\hbox{E}\kern-.125emX}}
\begin{document}

\title{Non-Orthogonal Multiple Access and Network Slicing: Scalable Coexistence of eMBB and URLLC}

\author{Eduardo Noboro Tominaga\IEEEauthorrefmark{1}, Hirley Alves\IEEEauthorrefmark{1}, Richard Demo Souza\IEEEauthorrefmark{2}, Jo\~{a}o Luiz Rebelatto\IEEEauthorrefmark{3}, Matti Latva-aho\IEEEauthorrefmark{1}\\

	\IEEEauthorblockA{
		\IEEEauthorrefmark{1}6G Flagship, Centre for Wireless Communications (CWC), University of Oulu, Finland\\
		\{eduardo.noborotominaga, hirley.alves, matti.latva-aho\}@oulu.fi\\
		\IEEEauthorrefmark{2}Federal University of Santa Catarina (UFSC), Florian\'{o}polis, Brazil, richard.demo@ufsc.br\\
		\IEEEauthorrefmark{3}Federal University of Technology - Paran\'{a} (UTFPR), Curitiba, Brazil, jlrebelatto@utfpr.edu.br
	}
}

\maketitle

\begin{abstract}
The 5G systems will feature three generic services: enhanced Mobile BroadBand (eMBB), massive Machine-Type Communications (mMTC) and Ultra-Reliable and Low-Latency Communications (URLLC). The diverse requirements of these services in terms of data-rates, number of connected devices, latency and reliability can lead to a sub-optimal use of the 5G network, thus network slicing is proposed as a solution that creates customized slices of the network specifically designed to meet the requirements of each service. Under the network slicing, the radio resources can be shared in orthogonal and non-orthogonal schemes. Motivated by Industrial Internet of Things (IIoT) scenarios where a large number of sensors may require connectivity with stringent requirements of latency and reliability, we propose the use of Non-Orthogonal Multiple Access (NOMA) to improve the number of URLLC users that are connected in the uplink to the same base station (BS), for both orthogonal and non-orthogonal network slicing with eMBB users. The multiple URLLC users transmit simultaneously and across multiple frequency channels. We set the reliability requirements for the two services and analyze their pair of sum rates. We show that, even with overlapping transmissions from multiple eMBB and URLLC users, the use of NOMA techniques allows us to guarantee the reliability requirements for both services.
\end{abstract}


\section{Introduction}

\par The fifth generation (5G) of wireless communication systems is currently under standardization and deployment around the world, and introduces three generic services: enhanced Mobile Broadband (eMBB), massive Machine-Type-Communications (mMTC) and Ultra-Reliable Low-Latency Communications (URLLC). eMBB aims to provide increased data rates, with peak rates on the order of gigabits per second to moderate rates in the order of megabits per second with high availability. The mMTC service aims at providing connectivity for a large number of cost- and energy-constrained devices (e.g. sensors) that often require low data rates. Finally, the challenging objective of URLLC is to provide ultra-reliable connectivity while operating in short block lengths, a requirement to achieve low latency demanded by time-critical applications \cite{metis5G}.

\par The current 5G New Radio (NR) network is not capable yet to satisfy the very stringent requirements of reliability and latency required by URLLC applications. That is one of the reasons why the research community has already started the development of solutions for beyond-5G and 6G wireless communications systems. One of the predicted use cases for beyond-5G and 6G systems is the critical wireless factory automation that requires communication with ultra-high reliability and ultra-low latency. Moreover, it is foreseen that the number of connected devices will increase substantially for 6G, which also poses very stringent requirements in terms of spectral efficiency \cite{6G_White_Paper}.In this context, 6G will scale the traditional URLLC to the massive connectivity dimension, thus leading to a new service class defined massive URLLC (mURLLC), that is the merge of the traditional URLLC and mMTC services from 5G \cite{saad2019}.

\par The diverse and sometimes conflicting requirements of the different 5G services and applications can lead to a sub-optimal use of the mobile network. One efficient solution is the slicing of the network in multiple virtual and isolated logical networks running on a common physical infrastructure in an efficient and economic way,
thus allowing slices to be individually customized with respect to, e.g., latency, energy efficiency, mobility, massive connectivity and throughput \cite{gsma}. Instead of having the traditional categorization into eMBB, mMTC and URLLC, some 6G applications will require dynamic service types, thus requiring the dynamic provisioning of network slices according to the data traffic and usage patterns of the MTC network \cite{nurul2020}.

\par The previous generations of wireless communications systems were mostly based on the utilization of Orthogonal Multiple Access (OMA) schemes to provide connectivity to multiple users. In such schemes, the users are allocated with radio resources that are orthogonal in time, frequency or code domain, and ideally no interference exists among them. However, one drawback of OMA schemes is that the maximum number of users is limited by the total amount of available orthogonal resources \cite{NOMA1}. To meet the diverse requirements of very high data rates, ultra-reliability, low latency, massive connectivity and spectral efficiency, Non-Orthogonal Multiple Access (NOMA) emerges as a promising technology for beyond-5G and 6G. NOMA allows multiple users to share time and frequency resources in the same spatial layer via power domain or code domain multiplexing \cite{NOMA1}. 
It is also predicted that in 6G scenarios there will be a need to new access schemes that can dynamically change between orthogonal and non-orthogonal multiple access schemes depending on the current state of the network \cite{saad2019}.


\par A communication-theoretic framework for network slicing in 5G was presented in \cite{popovski2018}, where the same radio resources are sliced among the heterogeneous 5G services under both orthogonal and non-orthogonal strategies. According to their definition of network slicing, under the orthogonal slicing for eMBB and URLLC, some frequency channels are allocated exclusively for the eMBB, while other frequency channels are allocated exclusively for the URLLC. On the other hand, under the non-orthogonal slicing, the same frequency channels can be shared by eMBB and URLLC services. In both schemes, the URLLC transmissions were allowed to span across multiple frequency channels as a way to explore diversity to achieve very stringent reliability requirements. However, they did not considered the use of NOMA for multiple URLLC users, such that the maximum number of URLLC users connected to the same Base-Station (BS) was limited by the number of minislots within the timeslot.

 \par The coexistence of eMBB and URLLC services has also been studied in the other works. Joint scheduling of eMBB and URLLC traffic has been studied in, for example, \cite{eMBB_URLLC_1}, \cite{eMBB_URLLC_2} and \cite{eMBB_URLLC_3}. Abreu \textit{et. al.} \cite{abreu2019} studied the multiplexing of eMBB and URLLC traffics in the uplink using an analytical framework. They considered the cases where different bands are allocated for each service, and also the case where both services share the same band. The slicing of resources for eMBB and URLLC has been also studied in \cite{eMBB_URLLC_4}, where the authors proposed a risk-sensitive based formulation to allocate resources to URLLC users while minimizing the risk of eMBB (i.e. protecting the eMBB users with low data rate) and ensuring URLLC reliability. In \cite{eMBB_URLLC_5} the authors adopted a time/frequency resource blocks approach to address the sum rate maximization problem subject to latency and slicing isolation constraints while guaranteeing the reliability requirements with the use of adaptive modulation coding. In \cite{eMBB_URLLC_6} the authors analyze the coexistence of eMBB and URLLC in fog-radio architectures where the URLLC traffic is processed at the edge while eMBB traffic is handled at the cloud. In \cite{eMBB_URLLC_7} the authors also study the orthogonal and non-orthogonal slicing of radio resources for eMBB and URLLC using a max-matching diversity (MMD) algorithm to allocate the frequency channels for the eMBB users. However, none of the mentioned works studied the performance of the joint combination of NOMA, SIC decoding and frequency diversity for URLLC traffic.

\par Motivated by the mURLLC scenarios predicted for beyond-5G and 6G networks, and based on recent works that address the coexistence between eMBB and URLLC (and in special \cite{popovski2018}), our contribution is the development a framework that allows multiple URLLC users to share the same radio resources with eMBB users in a scalable manner, for both orthogonal and non-orthogonal slicing of radio resources in the uplink scenario. The main difference of our work in comparison with \cite{popovski2018} is that we consider the use of NOMA for multiple URLLC devices. To achieve this goal, we propose an innovative approach based on the joint use of NOMA, Successive Interference Cancellation (SIC) and frequency diversity as a solution to improve the number of URLLC devices that can be connected to the same BS. In other words, we allow more than one URLLC user to be active in the same time/frequency resource and, in order to recover the packets from the multiple URLLC users and from the eMBB users, the BS performs SIC decoding. To characterize the performance trade-offs between eMBB and URLLC, we evaluate the pair of achievable sum rates under pre-defined reliability requirements for orthogonal and non-orthogonal scenarios. We show that, even with overlapping transmissions from multiple URLLC users, the use of NOMA SIC and frequency diversity techniques allow us to guarantee the reliability requirements of eMBB and URLLC services in both slicing schemes.

\par This paper is organized as follows. In the next section we present the system model and the individual performance analysis of the eMBB and URLLC services. In Section \ref{slicing for embb and urllc} we show how eMBB and URLLC users can share the same radio resources for both orthogonal and non-orthogonal slicing. In Section \ref{numerical illustration} we present the numerical results illustrating the performance trade-off between the services. Finally, the conclusions are presented in Section \ref{conclusions}.

\section{System Model}
\label{system model}

\par
\begin{figure}[!b]
	\centering
	\includegraphics[scale=0.75]{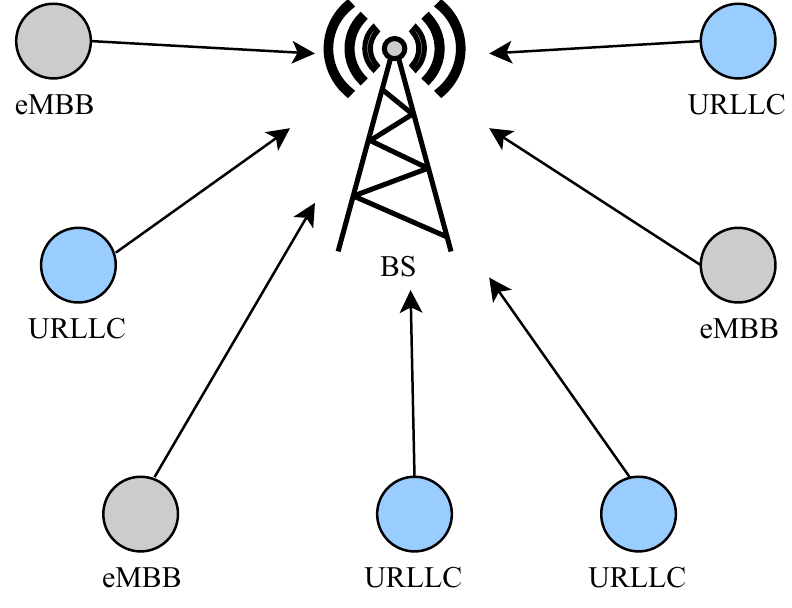}
	\caption{Uplink transmissions to a common base station (BS) from multiple eMBB and URLLC users.}
	\label{scenario}
\end{figure}

 We consider the uplink of a beyond-5G/6G network, where multiple eMBB and URLLC devices aim at transmitting independent packets to a common BS, as illustrated in Fig. \ref{scenario}. As in \cite{popovski2018}, we also consider a time-frequency grid composed of $F$ frequency channels indexed by $f \in \left\{1,\ldots,F\right\}$ and $S$ minislots indexed by $s \in \left\{1,\ldots,S\right\}$. The set of $S$ minislots composes a timeslot.

\par
\begin{figure}[h]
	\centering
	\includegraphics[scale=0.8]{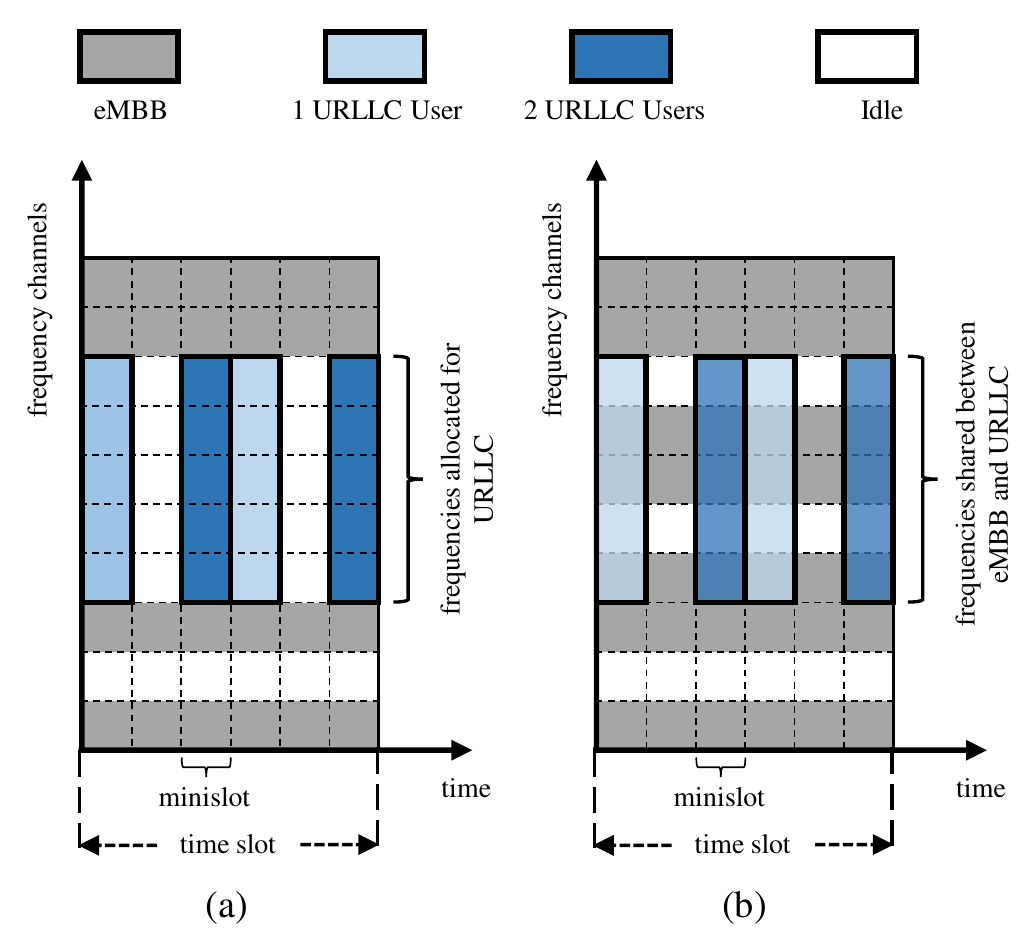}
	\caption{Illustration of the time-frequency grid used for the network slicing for the eMBB and URLLC services in the (a) orthogonal and (b) non-orthogonal scenarios. The darker blue tone indicates the overlap of URLLC transmissions.}
	\label{slicing}
\end{figure}

 The orthogonal and non-orthogonal slicing of the radio resources for eMBB and URLLC are based on \cite{popovski2018} and illustrated in Fig. \ref{slicing}, for $F = 10$ frequency channels, from which $F_U = 5$ frequency channels allocated for the URLLC traffic, and $S = 6$ minislots. However, differently from \cite{popovski2018}, we allow more than one URLLC user to transmit in the same minislot, as indicated by the darker blue tone in Fig. \ref{slicing}.

\par The transmission of an eMBB user occupies a single frequency channel $f$ and extends over the entire timeslot. Moreover, for eMBB traffic, we model only the standard scheduled transmission phase, hence assuming that radio access and competition among eMBB users have been solved prior to the considered time slot. A URLLC user, in turn, transmits within a single minislot across a  subset of $F_U \leq F$ frequency channels, as a mean to achieve frequency diversity and meet the reliability requirements \cite{popovski2018}. Due to the low latency requirement, each URLLC packet must be decoded within the duration of a minislot, not being allowed to span over multiple minislots. We assume that a massive number of URLLC users may be connected to the same BS, but only a subset of them are active simultaneously. Each radio resource $f$ is assumed to be within the time- and frequency-coherence interval of the wireless channel, so that the wireless channel coefficients are constant within each minislot. We also assume that the coefficients fade independently across the radio resources. The channel envelops as seen by the eMBB and the URLLC traffics, which we denote by $H_{i,f}$ with $i \in \left\{B,U\right\}$, are independent and complex Gaussian distributed, i.e., $H_{i,f} \sim \mathcal{CN}(0,\Gamma_i)$, thus the channel gains $G_{i,f} = |H_{i,f}|^2$ are exponentially distributed with average $\Gamma_i$. Moreover, no Channel-State Information (CSI) is assumed at the URLLC devices, whereas the eMBB devices and BS are assumed to have perfect CSI as in \cite{popovski2018}. The outage probabilities of the eMBB and URLLC devices are denoted as $\Pr(E_B)$ and $\Pr(E_U)$, respectively, and must satisfy the reliability requirements $\Pr(E_B) \leq \epsilon_B$ and $\Pr(E_U) \leq \epsilon_U$.

\subsection{Signal Model}

\par We define the maximum number of URLLC devices transmitting simultaneously as $n_U \in \left\{ \mathbb{N}^+\right\}$. The signal vector received at the BS in minislot $s$ and frequency channel $f$ is
\begin{equation}
    \label{eq1}
    \textit{Y}_{s,f} = H_{B,f}\textbf{X}_{B,s,f} + \sum_{u=1}^{n_U}H_{U_{u},f}\textbf{X}_{U_{u},s,f} + \textbf{Z}_{s,f},
\end{equation}
where $\textbf{X}_{B,s,f}$ is the signal transmitted by an eMBB user scheduled in the frequency channel $f$, $\textbf{X}_{U_u,s,f}$ is the signal transmitted by the active URLLC device of index $u$ in frequency channel $f$, and $\textbf{Z}_{s,f}$ is the noise vector, whose entries are i.i.d. Gaussian with zero mean and unit variance.

\par Under orthogonal slicing some frequency channels are allocated exclusively for the eMBB traffic and some exclusively for the URLLC traffic, whereas with non-orthogonal slicing the frequency channels can be shared between the two services. Besides, in this work we evaluate the performance of a worst case scenario, where there is always an eMBB user transmitting in each frequency channel $f$ and the $n_U$ URLLC users active in all the minislots. 

\subsection{Performance Analysis of the eMBB User}

\par Following \cite{popovski2018},  the objective is to transmit at the largest rate $r_{B,f}$ that is compatible with the outage probability requirement $\epsilon_B$ under a long-term average power constraint\footnote{Notice that full CSI of eMBB user is assumed as in \cite{popovski2018}. Since eMBB transmissions are scheduled, devices have sufficient time to undergo through CSI acquisition procedures \cite{popovski2018,eMBB_URLLC_3}.}. This can be formulated as the optimization problem \cite{popovski2018}
\begin{equation}
    \label{optimal rb}
	\begin{aligned}
		\text{maximize }&r_{B,f} \\
		\text{subject to }&\text{Pr}[\text{log}_2(1 + G_{B,f}P_B(G_{B,f}))\leq r_{B,f}] \leq \epsilon_B \\
		\text{and }&\mathbb{E}[P_B(G_{B,f})] = 1,
	\end{aligned}	
\end{equation}
where $P_B(G_{B,f})$ is the instantaneous transmit power. The optimal solution to this problem is given by the power inversion scheme. The eMBB device chooses a transmission power that is inversely proportional to the channel gain $G_{B,f}$ if the latter is above a given threshold $G_{B,f}^{\min}$, while it refrains from transmitting otherwise \cite{popovski2018}.

\par In the absence of interference from other services, the only source of outage for an eMBB transmission is the event that an eMBB device does not transmit because of insufficient SNR. The probability that $G_{B,f}$ is below $G_{B,f}^{\min}$ is \cite{popovski2018}
\begin{equation}
	\Pr(E_B) = \Pr[G_{B,f} < G_{B,f}^{\min}] = 1 - \exp\left(-\dfrac{G_{B,f}^{\min}}{\Gamma_B}\right).
\end{equation}

\par Imposing the reliability requirement $\Pr(E_B) = \epsilon_B$, the threshold SNR becomes
\begin{equation}
\label{Gb_min}
	G_{B,f}^{\min}=\Gamma_B\ln\left(\dfrac{1}{1-\epsilon_B}\right).
\end{equation}
Based on the power inversion scheme, the instantaneous power $P_B(G_{B,f})$ chosen as a function of the channel gain $G_{B,f}$ is
\begin{equation}
	P_B(G_{B,f}) = 	\begin{cases}
						\dfrac{G_{B,f}^{\text{tar}}}{G_{B,f}} &\text{if } G_{B,f} \geq G_{B,f}^{\min} \\
						0 &\text{if } G_{B,f} < G_{B,f}^{\min},
					\end{cases}
\end{equation}
where $G_{B,f}^{\text{tar}}$ is the target SNR, which is obtained by imposing the average power constraint as \cite{popovski2018}:

$$\mathbb{E}[P_B(G_{B,f})] = \dfrac{G_{B,f}^{\text{tar}}}{\Gamma_B} \Gamma \left(0,\tfrac{G_{B,f}^{\min}}{\Gamma_B}\right) = 1.$$

\par This implies that the target SNR is
\begin{equation}
    \label{Gb_tar}
    G_{B,f}^{\text{tar}} = \dfrac{\Gamma_B}{\Gamma \left(0,\tfrac{G_{B,f}^{\min}}{\Gamma_B}\right)},
\end{equation}
where $\Gamma(a,z)= \int_{z}^{\infty} t^{a-1}e^{-t}\mathrm{d}t$ is the upper incomplete gamma function. Finally, the outage rate achieved by the eMBB device is \cite{popovski2018}
\begin{equation}
    \label{rb_orth}
	r_B^{\text{orth}} = \log_2(1 + G_{B,f}^{\text{tar}}).
\end{equation}

\subsection{Performance Analysis of the URLLC User}
\label{IIC}

\par The URLLC user transmits data across $F_U$ frequency channels in a minislot. Due to the NOMA behavior of URLLC devices, there is always interference when $n_U > 1$. Since the URLLC devices are assumed to have no CSI and for mathematical tractability and simplicity, herein, we consider that all of them transmit with the same data rate $r_U$.

\par As mentioned previously, in this work we adopt a worst case assumption that there are always $n_U \in \left\{\mathbb{N}^+\right\}$ URLLC devices transmitting in all minislots. The BS performs SIC\footnote{Note that, as presented in \cite{popovski2018}, SIC outperforms other techniques of multi-user detection, such as puncturing.} decoding to recover the multiple packets that arrive at the same minislot. Let $u$ denote the index of an active URLLC device with channel gain $G_{U_u,f}$ for the allocated frequency channel $f$. Let us denote a SIC decoding ordering $\left\{1,\cdots,n_N\right\}$. The Signal-to-Interference-plus-Noise Ratio (SINR) in the frequency channel $f$ when decoding the $u$-th active URLLC device reads
\begin{equation}
    \sigma_{u,f} = \dfrac{G_{U_u,f}}{1+\sum\limits_{\substack{j>u}}^{n_U} G_{U_j,f}}.
\end{equation}

\par The active URLLC device is decoded successfully if
\begin{equation}
    \dfrac{1}{F_U}\sum_{f=1}^{F_U}\text{log}_2(1 + \sigma_{u,f}) \geq r_U.
\end{equation}

\par During the SIC decoding procedure, the BS first attempts to decode the strongest user among all the active URLLC users in a minislot. If this user is correctly decoded, its interference is subtracted from the received signal, then the BS attempts to decode the next user in the order of strongest users, and so on. The SIC decoding procedure ends when the decoding of one active URLLC user fails or after all the active URLLC users have been correctly decoded. We assume that all SIC decoding steps can be realized within the time duration of a minislot.

\par For simulation purposes, and in order to emulate the behaviour of the BS while performing the detection and SIC decoding of the URLLC users, we define the SIC decoding ordering according to their sum of mutual information across the $F_U$ frequency channels, which is defined by
\begin{equation}
    I_u^{\text{sum}} = \sum_{f=1}^{F_U}\text{log}_2(1 + \sigma_{u,f}).
\end{equation}

\par Given the reliability condition $\text{Pr}(E_U) \leq \epsilon_U$, the objective is to obtain the maximum rate $r_U$ that is a function of the number of frequency channels allocated for URLLC traffic. The URLLC sum rate, that is, the sum of the data rates of the $n_U$ active URLLC devices transmitting in a minislot, is given by
\begin{equation}
    r_{U}^{\text{sum}} = n_U \, r_U.
\end{equation}

\par Increasing $F_U$ enhances the frequency diversity and, hence, makes it possible to satisfy the reliability target $\epsilon_U$ at a larger rate $r_U$ \cite{popovski2018}. Besides,  during the computation of $r_U$, the error probabilities for all URLLC users are computed individually.

\section{Slicing for eMBB and URLLC}
\label{slicing for embb and urllc}

\par In this section, we consider the coexistence of eMBB and NOMA URLLC devices for both orthogonal and non-orthogonal slicing.

\subsection{Orthogonal Slicing between eMBB and URLLC}

\par Under the orthogonal slicing scenario, $F_U$ out of $F$ frequency channels for all minislots are allocated for URLLC traffic, while the remaining $F_B = F - F_U$ channels are each allocated to one eMBB user. The performance of the system is specified in terms of the pair $(r_B^{\text{sum}},r_U^{\text{sum}})$ of eMBB sum-rate $r_B^{\text{sum}}$ and URLLC sum-rate $r_U^{\text{sum}}$.

\par The eMBB sum-rate is obtained by \cite{popovski2018}
\begin{equation}
    r_B^{\text{sum}} = (F - F_U)r_B^{\text{orth}},
\end{equation}
where $r_B^{\text{orth}}$ is given by (\ref{rb_orth}).
Given a number $F_U$ of frequency channels allocated for URLLC, we compute the maximum URLLC rate $r_U$ that guarantees the reliability constraint $\text{Pr}(E_U) \leq \epsilon_U$ for all $n_U$ URLLC devices transmitting simultaneously, as detailed in Section \ref{IIC}.

\subsection{Non-Orthogonal Slicing between eMBB and URLLC}

\par In the non-orthogonal slicing scenario, all $F$ frequency channels are used for both eMBB and URLLC services. Hence, $F_U = F_B = F$. Due to the latency constraints, the decoding of a URLLC transmission cannot wait for the decoding of eMBB traffic. The eMBB requirements are less demanding in terms of latency, and hence eMBB decoding can wait for the URLLC transmissions to be decoded first. This enables a SIC mechanism whereby URLLC packets are successive decoded and then canceled from the received signal prior to the decoding of the eMBB signal \cite{popovski2018}. As a consequence, during the decoding attempts of the URLLC packets, the interference of the eMBB traffic is always present.

\par In the orthogonal case, as shown in (\ref{Gb_tar}), the variable $G_{B,f}^{\text{tar}}$ is uniquely determined by the error probability target $\epsilon_B$ and the threshold SNR $G_{B,f}^{\min}$ defined in (\ref{Gb_min}). For the non-orthogonal slicing, it may be beneficial to choose a smaller target SNR than the one given if (\ref{Gb_tar}), so as to reduce the interference caused to URLLC transmissions. This yields the inequality~\cite{popovski2018}
\begin{equation}
    \label{Gb_tar_max}
    G_{B,f}^{\text{tar}} \leq \frac{\Gamma_B}{\Gamma \left(0,\tfrac{G_{B,f}^{\min}}{\Gamma_B}\right)}.
\end{equation}

\par The maximum allowed SNR for the eMBB devices, which we denote by $G_{B,\max}^{\text{tar}}$, is set by the inequality in (\ref{Gb_tar_max}). Consequently, the maximum allowed eMBB data rate is given by $r_B^{\max} = \text{log}_2(1 + G_{B,\max}^{\text{tar}})$ .

\par The objective of the analysis is to determine the rate pair $(r_B^{\text{sum}},r_U^{\text{sum}})$ for which the reliability requirements of the two services are satisfied. To this end, first we fix an eMBB data rate $r_B \in [0,r_B^{\max}]$ and then we compute the maximum achievable rate $r_U$. During this computation, for a given value of $r_U$, we search for the minimum value of the SNR $G_B^{\text{tar}} \in [G_B^{\min},G_{B,\max}^{\text{tar}}]$ that can be used for all eMBB devices. The error probabilities for all eMBB and URLLC users are computed individually.  

\par Let us again define a SIC decoding ordering $\left\{1,\cdots,n_U\right\}$. The SINR in the frequency channel $f$ while decoding the $u$-th active URLLC device is given by
\begin{equation}
    \label{eq16}
    \sigma_{u,f} = \dfrac{G_{U_u,f}}{1+G_{B,f}^{\text{tar}}+\sum\limits_{\substack{j>u}}^{n_U} G_{U_j,f}}.
\end{equation}

\par The URLLC device with index $u$ is correctly decoded if
\begin{equation}
    \dfrac{1}{F_U}\sum_{f=1}^{F_U}\text{log}_2(1+\sigma_{u,f}) \geq r_U,
\end{equation}
where $\sigma_{u,f}$ is given by (\ref{eq16}).

\par The SIC decoding performed in the non-orthogonal slicing is similar to the procedure performed in the orthogonal scenario and described in Section \ref{IIC}. The only difference is that in the non-orthogonal slicing there is always interference from the eMBB users in all $F_U$ frequency channels during the SIC decoding of the URLLC users. Moreover, in given a minislot, the BS tries to decode the packets from the eMBB users only after all the URLLC users have been correctly decoded. If an error occurs during the SIC decoding of the URLLC users, the packets from the eMBB users are lost. 

\section{Numerical Results}
\label{numerical illustration}

\par In this section we present Monte Carlo simulation results for the orthogonal and non-orthogonal slicing of radio resources between eMBB and URLLC. For the sake of tractability, we consider the cases where the maximum number of URLLC devices transmitting simultaneously in the same minislot is $n_U \in \left\{1,2,3,4\right\}$, where $n_U=1$ is the scenario from \cite{popovski2018}. Besides, we consider a time-frequency grid with $F = 10$ frequency channels and reliability requirements $\epsilon_U = 10^{-5}$ and $\epsilon_B = 10^{-3}$ \cite{popovski2018}. Besides, we assume that the URLLC and eMBB devices are located in a similar environment, e.g. within a sector of a factory floor, aggregated by their proximity to the BS. This renders the same average received SNR to all the devices belonging to the same service class. 

\begin{figure}[t]
	\centering
	\includegraphics[scale=0.45]{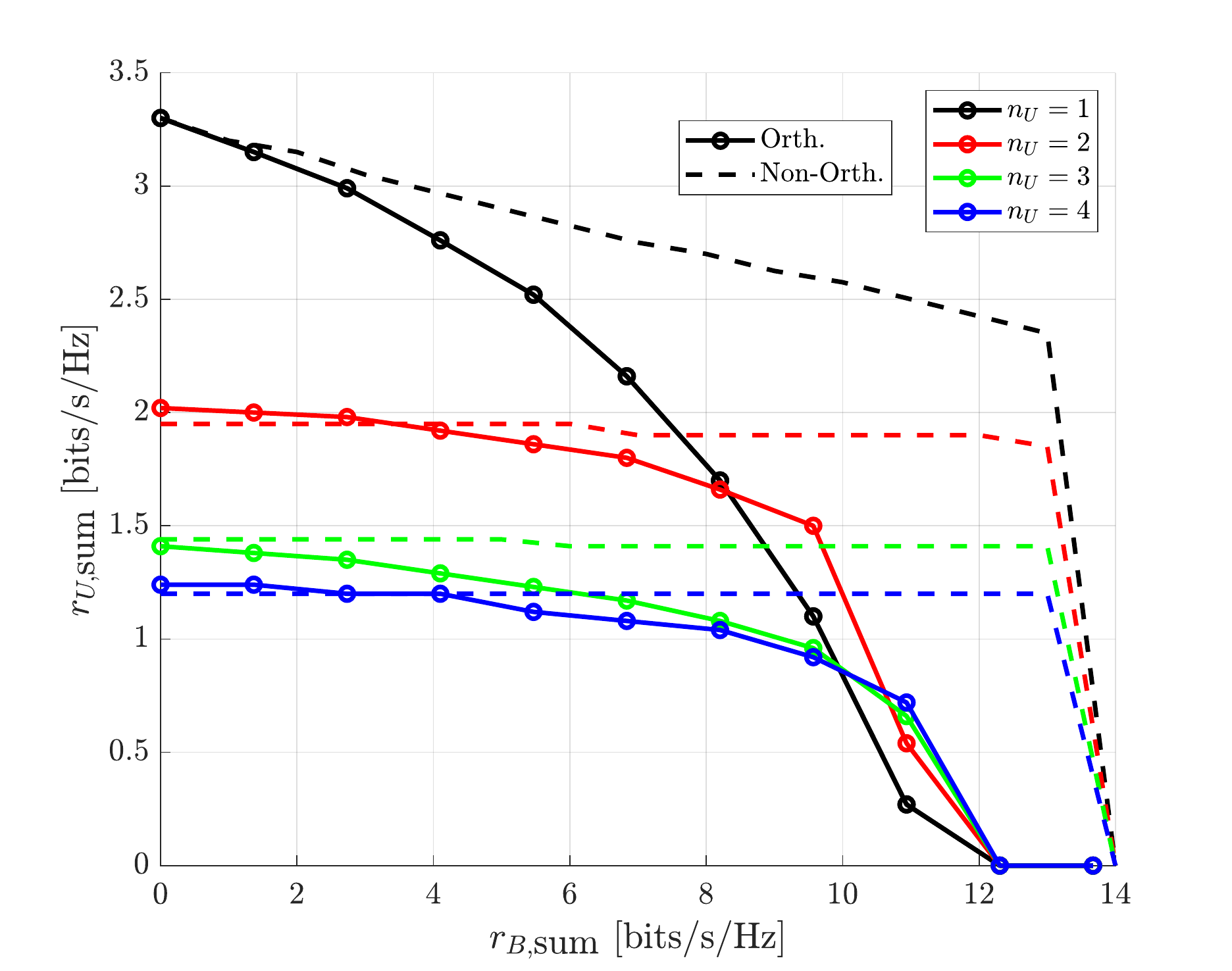}
	\caption{eMBB sum rate $r_{B}^{\text{sum}}$ versus URLLC sum rate $r_{U}^{\text{sum}}$ for the the orthogonal and non-orthogonal slicing when $\Gamma_U = 20 \text{ dB}$, $\Gamma_B = 10 \text{ dB}$, $F=10$, $\epsilon_U = 10^{-5}$ and $\epsilon_B = 10^{-3}$.}
	\label{result1}
\end{figure}

\begin{figure}[h!]
	\centering
	\includegraphics[scale=0.45]{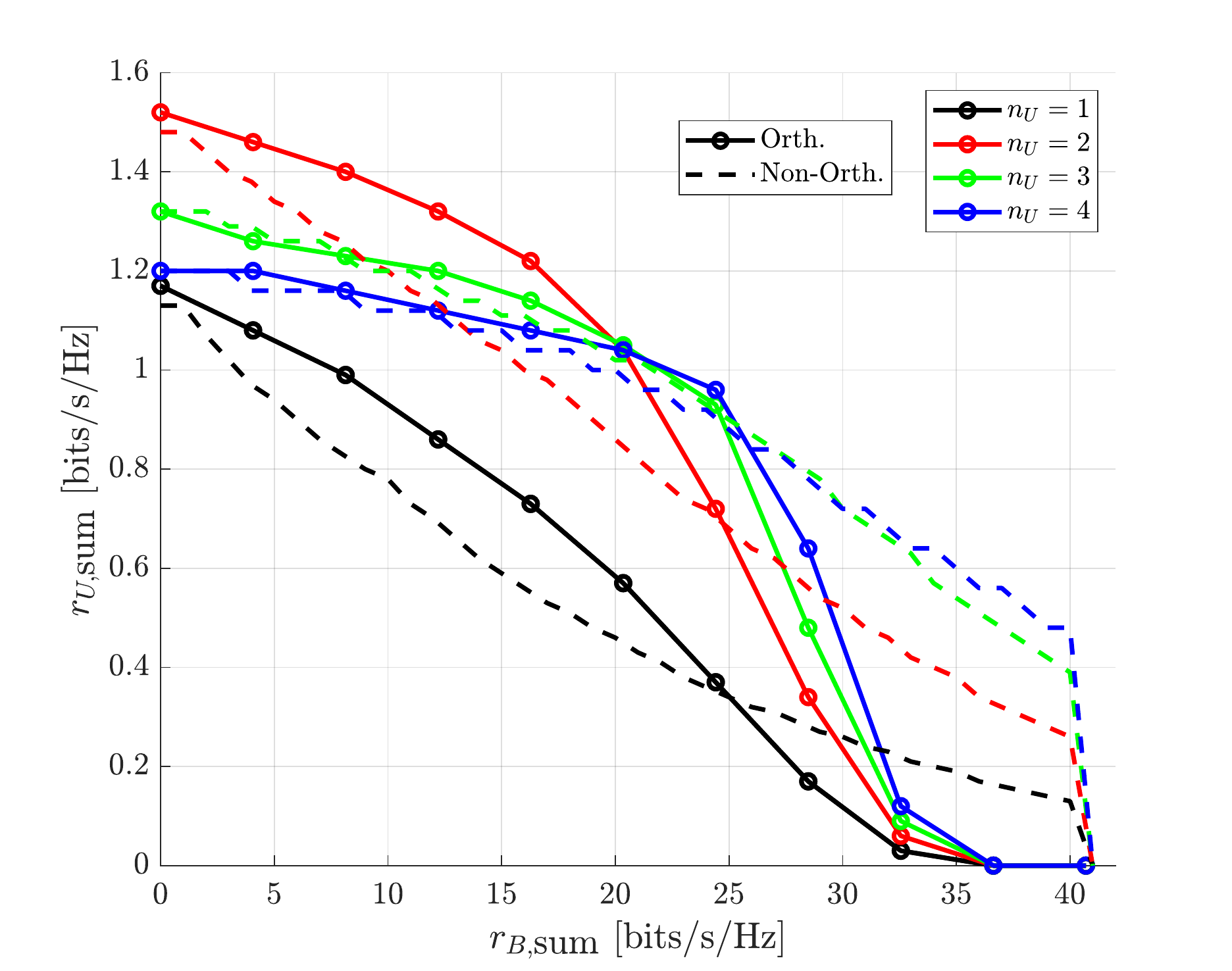}
	\caption{eMBB sum rate $r_{B}^{\text{sum}}$ versus URLLC sum rate $r_{U}^{\text{sum}}$ for the the orthogonal and non-orthogonal slicing when $\Gamma_U = 10 \text{ dB}$, $\Gamma_B = 20 \text{ dB}$, $F=10$, $\epsilon_U = 10^{-5}$ and $\epsilon_B = 10^{-3}$.}
	\label{result2}
\end{figure}

\par We plot the pair of sum rates for the orthogonal and non-orthogonal slicing scenarios considering average channel gains of $\Gamma_U = 20 \text{ dB}$ and $\Gamma_B = 10 \text{ dB}$ in Fig. \ref{result1} and $\Gamma_U = 10 \text{ dB}$ and $\Gamma_B = 20 \text{ dB}$ in Fig. \ref{result2}. For $\Gamma_U > \Gamma_B$, the highest values of $r_U^{\text{sum}}$ are achieved when only one URLLC user is active in the minislot, that is, $n_U=1$, as showed by the black curves in Fig. \ref{result1}. As we increase $r_B^{\text{sum}}$, the non-orthogonal slicing allows us to achieve pairs of sum rates that are not possible to achieve using the orthogonal slicing, as illustrated by the dashed curves in Fig. \ref{result1}. For both orthogonal and non-orthogonal slicing and $\Gamma_U>\Gamma_B$, the $r_U^{\text{sum}}$ decreases as we increase $n_U$, but interestingly, in the setups with $n_U>1$, the values of $r_U^{\text{sum}}$ do not vary much in the orthogonal slicing even for high increases of $r_B^{\text{sum}}$, while the non-orthogonal slicing makes them be almost constant for the whole range of $r_B^{\text{sum}}$.

\par In Fig. \ref{result2} we consider an opposite scenario where $\Gamma_U<\Gamma_B$. Differently from the case where $\Gamma_U>\Gamma_B$, now the setups with $n_U>1$ outperform the case with $n_U=1$, as showed by the achieved pairs of sum rates in Fig. \ref{result2}. When $\Gamma_U<\Gamma_B$, it is possible to achieve higher values of $r_B^{\text{sum}}$, but at the cost of lower values of $r_U^{\text{sum}}$. For a large range of $r_B^{\text{sum}}$, the orthogonal slicing outperforms the non-orthogonal slicing, while the non-orthogonal slicing outperforms the orthogonal slicing only for the highest values of $r_B^{\text{sum}}$, as illustrated by the dashed curves Fig. \ref{result2}. In both Figs. \ref{result1} and \ref{result2}, the curves obtained for $n_U=1$ match those in \cite{popovski2018}.

\par Overall, when $\Gamma_U>\Gamma_B$ it is possible to achieve higher URLLC sum rates at cost of lower eMBB sum rates. However, the URLLC sum rate reduces as the number of connected devices increases. Therefore, under these conditions and for applications that impose strict rate constraints, it is better to limit $n_{U}\leq 2$, which provides the largest gains. On the other hand, the situation where $\Gamma_U<\Gamma_B$ is favorable to eMBB traffic in terms of the achievable sum rates. However, under this condition the NOMA URLLC is advantageous, since it is possible to achieve higher sum rates when we have multiple URLLC users connected to the same BS\footnote{Overall, the results indicate that NOMA provides larger gains for limited number of users, in this case $n_U=2$, which corroborated by e.g. \cite{onelTWC2018}.}. We conclude that the non-orthogonal slicing is the best choice only for applications that require very high eMBB sum rates.

\section{Conclusions}
\label{conclusions}

\par We proposed the use of NOMA and SIC decoding as a solution to improve the number of URLLC users that are connected to the same BS and when they share the same radio resources with eMBB users, in both orthogonal and non-orthogonal slicing of radio resources between the two services. We showed that, using the proposed method, multiple eMBB and URLLC users can transmit overlapping packets to same BS while their reliability requirements are still met. We demonstrated through simulations that when the URLLC users have better channel conditions than the eMBB users, the non-orthogonal slicing is advantageous over the orthogonal slicing for the whole range of eMBB sum rates. However, when the eMBB users have better channel conditions than the URLLC users, the non-orthogonal slicing outperforms the orthogonal slicing only for very high values of eMBB sum rates, that is, the orthogonal slicing is the best option for a wide range of eMBB sum rates.

\par As stated in \cite{6G_White_Paper}, the current 5G NR network is not yet capable of meeting the very stringent requirements of URLLC in terms of latency and reliability, and meeting these requirements requires hyper-flexible networks where technologies such as Artificial Intelligence (AI) and machine learning can be use to determine the optimal radio resource allocations for BSs and users. The framework developed in this work can be used in the specification of such 6G networks for mURLLC scenarios, where a large number of devices used for the control and/or monitoring of critical processes may require URLLC connectivity in the coexistence with other applications that require the high data rates provided by eMBB.


\section*{Acknowledgment}

\par This research has been financially supported by Academy of Finland, 6Genesis Flagship (grant no 318927), FIREMAN (no 326301) and Aka Prof (no 307492).

\bibliographystyle{./bibliography/IEEEtran}
\bibliography{./bibliography/main}

\begin{thebibliography}{10}
\providecommand{\url}[1]{#1}
\csname url@samestyle\endcsname
\providecommand{\newblock}{\relax}
\providecommand{\bibinfo}[2]{#2}
\providecommand{\BIBentrySTDinterwordspacing}{\spaceskip=0pt\relax}
\providecommand{\BIBentryALTinterwordstretchfactor}{4}
\providecommand{\BIBentryALTinterwordspacing}{\spaceskip=\fontdimen2\font plus
\BIBentryALTinterwordstretchfactor\fontdimen3\font minus
  \fontdimen4\font\relax}
\providecommand{\BIBforeignlanguage}[2]{{%
\expandafter\ifx\csname l@#1\endcsname\relax
\typeout{** WARNING: IEEEtran.bst: No hyphenation pattern has been}%
\typeout{** loaded for the language `#1'. Using the pattern for}%
\typeout{** the default language instead.}%
\else
\language=\csname l@#1\endcsname
\fi
#2}}
\providecommand{\BIBdecl}{\relax}
\BIBdecl

\bibitem{metis5G}
H.~{Tullberg}, P.~{Popovski}, Z.~{Li}, M.~A. {Uusitalo}, A.~{Hoglund},
  O.~{Bulakci}, M.~{Fallgren}, and J.~F. {Monserrat}, ``{The METIS 5G System
  Concept: Meeting the 5G Requirements},'' \emph{IEEE Commun. Mag.}, vol.~54,
  no.~12, pp. 132--139, December 2016.

\bibitem{6G_White_Paper}
M.~{Latva-Aho} and K.~{Lepp\"{a}nen}, ``{Key Drivers and Research Challenges
  for 6G Ubiquitous Wireless Intelligence},'' in \emph{6G Wireless Summit,
  Levi, Finland}, Mar 2019.

\bibitem{saad2019}
W.~{Saad}, M.~{Bennis}, and M.~{Chen}, ``{A Vision of 6G Wireless Systems:
  Applications, Trends, Technologies, and Open Research Problems},'' \emph{IEEE
  Network}, vol.~34, no.~3, pp. 134--142, 2020.

\bibitem{gsma}
\BIBentryALTinterwordspacing
{GSM Association}, ``{An Introduction to Network Slicing},'' Tech. Rep., 2017.
  [Online]. Available:
  \url{https://www.gsma.com/futurenetworks/wp-content/uploads/2017/11/GSMA-An-Introduction-to-Network-Slicing.pdf}
\BIBentrySTDinterwordspacing

\bibitem{nurul2020}
N.~H. {Mahmood}, H.~{Alves}, O.~A. {López}, M.~{Shehab}, D.~P.~M. {Osorio},
  and M.~{Latva-Aho}, ``{Six Key Features of Machine Type Communication in
  6G},'' in \emph{2020 2nd 6G Wireless Summit (6G SUMMIT)}, 2020, pp. 1--5.

\bibitem{NOMA1}
L.~{Dai}, B.~{Wang}, Y.~{Yuan}, S.~{Han}, C.~{I}, and Z.~{Wang},
  ``{Non-Orthogonal Multiple Access for 5G: Solutions, Challenges,
  Opportunities, and Future Research Trends},'' \emph{IEEE Commun. Mag.},
  vol.~53, no.~9, pp. 74--81, Sep. 2015.

\bibitem{popovski2018}
P.~{Popovski}, K.~F. {Trillingsgaard}, O.~{Simeone}, and G.~{Durisi}, ``{5G
  Wireless Network Slicing for eMBB, URLLC, and mMTC: A Communication-Theoretic
  View},'' \emph{IEEE Access}, vol.~6, pp. 55\,765--55\,779, 2018.

\bibitem{eMBB_URLLC_1}
Z.~{Wu}, F.~{Zhao}, and X.~{Liu}, ``{Signal Space Diversity Aided Dynamic
  Multiplexing for eMBB and URLLC Traffics},'' in \emph{2017 ICCC}, Dec 2017,
  pp. 1396--1400.

\bibitem{eMBB_URLLC_2}
A.~{Anand}, G.~{De Veciana}, and S.~{Shakkottai}, ``{Joint Scheduling of URLLC
  and eMBB Traffic in 5G Wireless Networks},'' in \emph{IEEE INFOCOM 2018},
  April 2018, pp. 1970--1978.

\bibitem{eMBB_URLLC_3}
A.~A. {Esswie} and K.~I. {Pedersen}, ``{Opportunistic Spatial Preemptive
  Scheduling for URLLC and eMBB Coexistence in Multi-User 5G Networks},''
  \emph{IEEE Access}, vol.~6, pp. 38\,451--38\,463, 2018.

\bibitem{abreu2019}
R.~{Abreu}, T.~{Jacobsen}, G.~{Berardinelli}, K.~{Pedersen}, N.~H. {Mahmood},
  I.~Z. {Kovacs}, and P.~{Mogensen}, ``{On the Multiplexing of Broadband
  Traffic and Grant-Free Ultra-Reliable Communication in Uplink},'' in
  \emph{2019 IEEE 89th Vehicular Technology Conference (VTC2019-Spring)}, 2019,
  pp. 1--6.

\bibitem{eMBB_URLLC_4}
M.~{Alsenwi}, N.~H. {Tran}, M.~{Bennis}, A.~{Kumar Bairagi}, and C.~S. {Hong},
  ``{eMBB-URLLC Resource Slicing: A Risk-Sensitive Approach},'' \emph{IEEE
  Commun. Lett.}, vol.~23, no.~4, pp. 740--743, April 2019.

\bibitem{eMBB_URLLC_5}
P.~K. {Korrai}, E.~{Lagunas}, S.~K. {Sharma}, S.~{Chatzinotas}, and
  B.~{Ottersten}, ``{Slicing Based Resource Allocation for Multiplexing of eMBB
  and URLLC Services in 5G Wireless Networks},'' in \emph{2019 IEEE CAMAD},
  Sep. 2019, pp. 1--5.

\bibitem{eMBB_URLLC_6}
R.~{Kassab}, O.~{Simeone}, P.~{Popovski}, and T.~{Islam}, ``{Non-Orthogonal
  Multiplexing of Ultra-Reliable and Broadband Services in Fog-Radio
  Architectures},'' \emph{IEEE Access}, vol.~7, pp. 13\,035--13\,049, 2019.

\bibitem{eMBB_URLLC_7}
E.~J. {dos Santos}, R.~D. {Souza}, J.~L. {Rebelatto}, and H.~{Alves},
  ``{Network Slicing for URLLC and eMBB With Max-Matching Diversity Channel
  Allocation},'' \emph{IEEE Communications Letters}, vol.~24, no.~3, pp.
  658--661, 2020.

\bibitem{onelTWC2018}
O.~L. {Alcaraz L\'{o}pez}, H.~{Alves}, P.~H. {Juliano Nardelli}, and
  M.~{Latva-aho}, ``{Aggregation and Resource Scheduling in Machine-Type
  Communication Networks: A Stochastic Geometry Approach},'' \emph{IEEE Trans.
  on Wireless Commun.}, vol.~17, no.~7, pp. 4750--4765, July 2018.

\end{thebibliography}

\end{document}